# Resource allocation using SPA based on different cost functions in elastic optical networks

Mehdi Tarhani, Sanjib Sarkar, Morad Khosravi Eghbal and Mehdi Shadaram

**Abstract**: Routing, modulation and spectrum allocation in elastic optical networks is a problem aiming at increasing the capacity of the network. Many algorithms such as shortest path algorithm can be used as the routing section of this problem. The efficiency of these algorithms is partly based on how the cost on each link is defined. In this study, we consider several basic costs and compare their effects on the network capacity. In particular, the static costs and the dynamic costs are evaluated and compared; For dynamic scenarios, compared to static scenarios, at least one additional factor is added that, for instance, can be usage of the link. We further consider a new factor that is based on probability of accommodating the signal at a given time in any given link. The results show that, among them, the shortest path algorithm provides the least blocking probability when the cost is a combination of link length and the abovementioned possibility/ usage of the link.

## I) Introduction

Ever increasing need for higher traffic resulting from the emerging applications has motivated new research to increase the capacity of data networks [1] – [2]. Elastic Optical Networks has been shown to be a good alternative for its counterpart Wavelength Division Multiplexing (WDM) because of not only higher capacity but also its flexibility and suitability for variable datarate traffics, one of the characteristics of the new data networks [3]. Orthogonal frequency multiplexing and intelligent client node, adaptive transmission and flexible grids are the characteristics of the Elastic optical Networks [4]. This new networking method has its own constraints as well.

The authors are with the Department of Electrical and Computer Engineering, University of Texas at San Antonio, San Antonio, TX 78249 USA (e-mail: mehdi.tarhani@utsa.edu)

In particular, wavelength continuity constraint requires that no frequency conversion is allowed over the path connecting a given pair of source and destinations. Further, wavelength contiguity constraint requires that all frequency slots dedicated for signal transferring the data between a sender and a receiver must be adjacent [3]. These are the reasons that all the previously worked algorithms for the wavelength division multiplexing need to be reinvestigated and adjusted to be efficient in EONs [4].

Routing, modulation and spectrum allocation (RMSA) is the main problem in EON aiming at making the resource allocation very efficient toward higher capacity and less resource utilization [5]. To determine path between a source and a destination there can be two general approaches [5] – [7]. First, a path (or a group of paths) is determined in advance of the network operation and this will be the only path between the pair whenever a new demand arrives to de network requiring transferring the data between them. Contrary to this static approach, in the other group of routing algorithms, the algorithm dynamically searches and finds a route based on the current state of the network and hence different routes may be found and used in different times and in different stats of the network. For both approaches, one of the simple and efficient algorithms that is commonly used [8]–[10] is shortest path algorithm (SPT). This algorithm receives the graph of the network with weighted edges (the weights are the costs that will be defined on each link) and finds a route that has the least total cost. Therefore the route and hence the efficiency of SPT depends heavily on what factors have been considered to define the cost of the links.

In this paper we consider some of the basic factors in both static and dynamic scenarios and compare their effects on the capacity of the networks. Last not the least, we consider three basic functions (linear, quadratic and square root) to represent three general ways of merging the factors involved in defined the cost and evaluate their effect on the performance of the network.

The rest of the paper is organized as follows. Section II provides the network model considered in this work. Section III describes the costs that are supposed be evaluated. The results of the simulations are provided in sections IV and finally section V concludes the paper.

## II) Network Model

We assume that bidirectional fiber links connect the nodes of the network. Full range optical spectrum in each link is 180 continuous frequency slices in which the central wavelength spacing is 12.5 GHZ. Demands can arrive to the network at any time and the source, destination and its required bitrate are random and unknown prior to arrival of the demand. The requested datarate varies from 1 to 50 GB/s, from demand to demand but is fixed for any given demand during the serving time. The modulation techniques that we consider in this paper include Binary Phase Shift keying (BPSK), quadrature phase shift keying (QPSK), 8-quadrature amplitude modulation 8QAM, and 16QAM. The maximum datarates per frequency slice they can support and the maximum supported distance under which each modulation techniques can deliver signal with acceptable SNR are given [11] in table I.

**Table I. Reach( in km) and maximum datarates (Gb/s)**

|  | BPSK | QPSK | 8QAM | 16QAM |
|---|---|---|---|---|
| Datarate (Max.) | 12.5 | 25 | 37.5 | 50 |
| Distance (Max.) | 5000 | 2500 | 1250 | 625 |

Network is assumed to be transparent. In other word, optical bypass is carried out in all intermediary nodes. For each path the algorithm employs a modulation format that can provide the maximum possible spectral efficiency as long as its length does not exceed the maximum distance supported by the corresponding modulation technique. The number of required frequency slices, for instance, for a demand with 40 GB/s is 3, ($\frac{40}{25}+1>2$), if the modulation techniques QPSK is used (one slice is allocated for guard band).

## III) Examined Costs

1) *Link Length*: This is a static cost because the length of the link does not change with network operation and the cost will remain the same regardless of the traffic in the network. The selected path by this cost will have the shortest length (total fiber lengths) and therefore most efficient modulation technique can be applied which means least possible number of frequency slices will be required for the signal over the path. However, there will not be a guarantee if the considered path will have enough available continuous frequency slices because the link cost does not consider the utilization of the link.

2) *Unity:* This is another static cost in which cost of all links are equal to one. If the modulation level employed ever the path by this cost is the same as the one with Link Length as the cost and the path are not the same then the resource utilization in this path is less because the number of links involved are less. However, similar to the previous one, this cost does not incorporate any information about the current usage of the links.

3) *Distance/unity and usage of the link*: This cost is dynamic in that the cost of the link can change from demand to demand because the usage of the link can be different in different times. This cost tries to merge both efficiency in spectral usage and also the uniformity of the resource utilization.

4) *Distance/link and the probability of accommodating a signal in a link*: The second part of this cost tries to not only create uniformity of resource utilization but also find a path that actually has the maximum chance to accommodate the signal for the given demand. For instance, suppose that a frequency slot that is available to use is surrounded by two frequency slots that are already in use by previous demands. For a demand that requires a minimum of three slots this slot should not be considered as available because of the wavelength contiguity constraint. The probability in this cost is simply the fraction of the total frequency slots in the link that not only are available but also can be used for a particular demand.

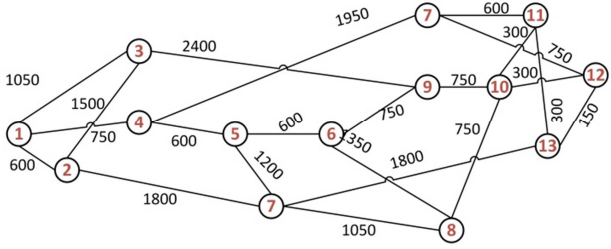
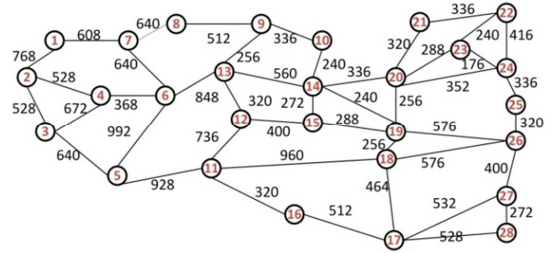

Fig. 1: Topologies,(left) NSFNet network, (right)US Backbone network

5) *Cost Functions*: The functions that merge the above mentioned metrics can also affect the performance of the networks. For the factors in section 3 we define cost $C$ as follows:

$$C = L + \alpha u^2 \qquad (1)$$

Where $L$ is the normalized length of the link, $u$ is the usage or the number of used slices over total number of slices in the link and $\alpha$ is a tanning coefficient. This quadratic term adds to the cost of the link especially when the link is about to be fully occupied the other two costs have linear and square root term to include the usage $u$ as follows:

$$C = L + \alpha u \qquad (2)$$

$$C = L + \alpha \sqrt{u} \qquad (3)$$

As we mentioned we will evaluate the effects of these three functions separately. When comparing the effect of the metrics we use $L$, 1, $L + \alpha u$, and $p + \alpha u$ ($p$ the possibility defined in section 4) respectively to consider the effects of metrics in section 1- 4.

IV) Simulation Results

In this section we evaluate the Shortest Path Algorithm for described cost functions in terms of blocking probability and per served demand transceiver usage. In the experiments, the distribution of the demands is randomly produced by a Poison distribution with $\lambda = 10$ and the duration of the demands follows an exponential distribution with variable $\mu$ to represent different traffic loads.

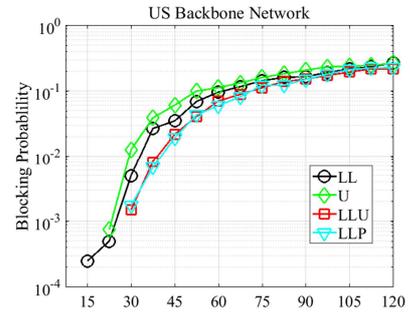

Fig. 2: Performance of SPT using different cost metrics

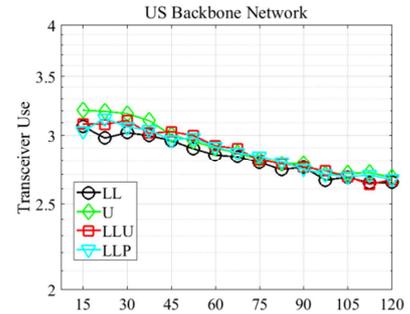

Fig. 3: Resource utilization of SPT using different cost metrics

The experiments are under two different networks namely US Backbone network, and NSFnet network Shown in Fig. 1. The results have been produced by using simulation using MATLAB. For each load in each cost we ran the network for 10000 demands to provide enough accuracy.

Fig. 2 and Fig. 3 show the blocking probability of the shortest path Algorithm using the metrics, length link (LL), unity (U), link length and link usage (LLU), and link length and probability of accommodating (LLP) as defined in section III. As it can be seen, the dynamic metrics (LLU and LLP) have less blocking probability compared to static ones (LL and U). Also, for static metrics, depending on the network, one of the metric is better than the other. Another observation is that the number of used transceivers per served demands in dynamic metrics is higher compared to static metrics which also implies that the route that is selected by them is not the one with the least resource requirements. In fact, the reason that they have better total performance is that they use the resources more uniformly across the network and, hence, there is a smaller probability of lacking enough adjacent frequency slots over a path for any particular demand.

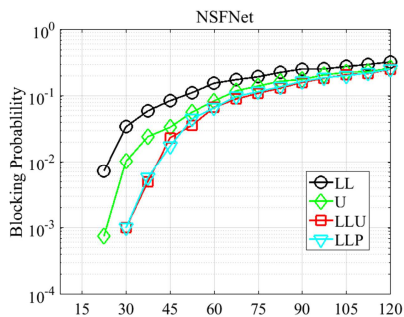

**Fig. 4: performance of SPT using different cost metrics**

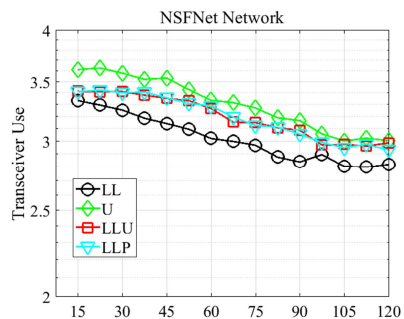

**Fig. 4 Resource utilization of SPT using different cost metrics**

Fig. 5 shows the blocking probability of the US backbone network under the cost defined by (1), (2), and (3) respectively. As we expected, the linear function has slightly less blocking probability the (3) has the slightly higher blocking probability.

## V) Conclusion

In this paper we examined the effect of different cost metrics on the blocking probability of two networks using the shortest path algorithm (SPT). The simulations results show that the dynamic metrics that take into account the resource utilization of the links can result in less total blocking probability although for any particular demand their selected path may not be the one with the least resource requirements.

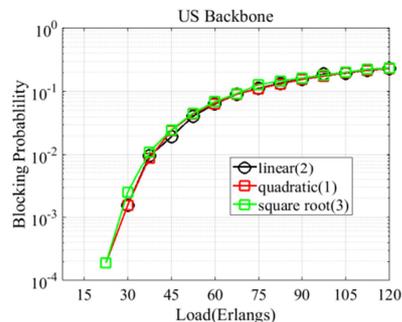

**Fig. 4: performance of SPT using (1) – (3)**